\documentclass[aps,prb,preprint,groupedaddress,showpacs]{revtex4}

\usepackage{graphicx} 

\begin{document}


\title{Zero Sound Attenuation near the Quantum Limit in Normal Liquid 
$^{3}$He close to the Superfluid Transition}


\author{Brian C. Watson}
\thanks{Present address: Information Systems 
Laboratories, Inc., San Diego, CA 92121, USA.}
\author{Naoto Masuhara}
\author{Mark W. Meisel}
\affiliation{Department of Physics and the Center for Condensed Matter Sciences, 
Microkelvin Research Laboratory, University of Florida, 
Gainesville, FL  32611-8440,  USA}


\date{\today}

\begin{abstract}
The zero sound attenuation of normal liquid $^{3}$He has been studied 
over a range of temperatures from slightly above the superfluid transition 
temperature, $T_{c}$, to approximately 10 mK at the constant pressures of 
1 bar and 5 bar.  Using longitudinal LiNbO$_{3}$ transducers, operating both 
on and off resonance, the experiment was performed at 15 discrete frequencies 
located in several broadband frequency windows, including $16-25$ MHz, 
$60-70$ MHz, and $105-111$ MHz.  
The results are compared to Landau's 
prediction for the attenuation of zero sound in the quantum limit, 
$(k_{B}T \ll \hbar \omega \ll k_{B}T_{F})$, where  
$\alpha_{0}(P,T,\omega) = 
\alpha^{\prime}(P)\,T^{2}\,\{1+(\hbar \omega/2 \pi k_{B}T)^{2}\}$.  
Calibration of the received zero sound signals was performed by 
measuring the temperature dependence of the first sound attenuation 
from 30 mK to 800 mK at those same frequencies and pressures.  The 
data are compared to previous results.
\end{abstract}


\maketitle

\section{INTRODUCTION}

In 1957, Landau established the foundations 
of a phenomenological theory used to describe strongly interacting 
Fermi systems.\cite{Landau1,Landau2}  Commonly referred to as 
Landau Fermi Liquid Theory, this theory has provided an extremely successful 
description of normal liquid $^{3}$He and some properties of superfluid $^{3}$He, 
and it has also been extended to 
electron systems in some metals.\cite{Nozieres,Pines,Leggett,Baym,VW,Dobbs,foot}  
Within the framework of this theory, quantum kinetic equations for Fermi 
liquids often reduce to classical kinetic equations, and the subtle 
differences between these two limits is not easily 
detected experimentally.  However, a significant difference is expected 
to appear in the attenuation of zero sound in normal liquid $^{3}$He when 
$2 \pi k_{B}T \simeq \hbar\omega$, where $\omega = 2 \pi f$ is the angular frequency 
of the sound excitation. 

In this paper, we provide a detailed description of our attempt to study  
the quantum limit of zero sound attenuation.  The next section reviews the 
nature of zero sound and the previous attempts to measure the attenuation in the 
quantum limit.  This introductory material is followed by an overview of  
our general approach to the experiment and the analysis of the results, 
especially the calibration necessary to obtain the absolute attenuation of 
the liquid.  The 
presentation continues with a description of our experimental cell and acoustic 
techniques.  Next, our data in the zero sound and first sound regimes will be 
presented, and then our analysis of data will be provided.  Finally, we will 
conclude with a discussion of the present state of our results.

\section{Review of Zero Sound and Attenuation Studies}

In his pioneering work, Landau also identified a new type of sound mode which is 
known 
as zero sound.  Distinct from ordinary hydrodynamic or first sound, zero sound 
occurs when the standard quasiparticle collisions can no longer provide the 
necessary restoring force to the system.  However, even in this ``collisionless'' 
regime, the quasiparticles experience restoring forces which attempt to return 
the system to equilibrium.  These forces arise from the interactions of the 
quasiparticles and enable the propagation of zero sound.  More specifically, the 
first sound regime exists when $\omega \tau \ll 1$ while zero sound propagates 
when $\omega \tau \gg 1$, where  
$\tau$ is the quasiparticle collision time which is 
theoretically predicted to be proportional to $T^{-2}$.  Consequently, by fixing 
the frequency of the probing sound wave and sweeping the temperature under the 
appropriate conditions, the theory predicts a crossover from zero sound to first 
sound when $\omega \tau = 1$.  The theoretical predictions for the attenuation of 
zero and first sound may be written, respectively, as
\begin{equation}
\alpha_{0}(P,T)\,=\,\alpha^{\prime}(P)\,T^{2} \;\;\; \textrm{for}\;\;\; 
\frac{1}{\tau} \ll \omega \;\;\;,
\end{equation}
and
\begin{equation}
\alpha_{1}(P,T,\omega)\,=\,\alpha^{\prime\prime}(P)\,
\left(\frac{\omega^{2}}{T^{2}}\right)\;\;\;
\textrm{for} \;\;\; \omega \ll \frac{1}{\tau} \;\;\;,
\end{equation}
where $\alpha^{\prime}$ and $\alpha^{\prime \prime}$ are appropriate 
pressure-dependent 
parameters.\cite{Baym}  In 1963, through the use of acoustic impedance techniques, 
Keen, Matthew, and Wilks experimentally verified the existence of a 
crossover from first to zero sound.\cite{Keen}  
The first direct observation of the first to zero 
sound regimes and the predicted behavior for the attenuation was reported in 1966 by 
Abel, Anderson, and Wheatley.\cite{Abel2}  Since this early work, numerous 
workers have 
measured $\alpha^{\prime}$, which is strongly pressure dependent.  The experiments 
reported herein have allowed us to determine $\alpha^{\prime}(P)$ at 1 bar and 
5 bar.  
In Appendix~A, our values are combined with the results of other 
groups,\cite{Lawson,Ketterson,Mast,Nara,Avenel} and 
interpolation formula are presented for $\alpha^{\prime}$ as a function of pressure 
and molar 
volume ranging from saturated melting pressure to the melting curve.

An additional important prediction of Landau Fermi Liquid Theory is the behavior of 
the zero sound attenuation in the limit 
$\hbar/\tau \ll k_{B}T \ll \hbar\omega \ll k_{B}T_{F}$.  
When $\hbar/\tau \ll \hbar\omega \ll k_{B}T$, 
the scattered quasiparticles remain within the thermally broadened Fermi surface.  
Due to the successive thermal collisions, the quantum mechanical features of the 
individual processes are smeared out.  In this case which 
Landau referred to as the ``classical'' limit,\cite{Landau2} the thermal 
collision rate simply determines the absorption of the ultrasound as expressed 
in Eq.~(1), and the sound attenuation vanishes if this condition is extrapolated to 
the zero temperature limit.
On the other hand when $\hbar\tau \ll k_{B}T \ll \hbar\omega \ll k_{B}T_{F}$, 
the quasiparticles are scattered away from the Fermi surface and remain in 
non-equilibrium states for a longer time due to the scarce collision probability.  
In this case which Landau referred to as the ``quantum'' regime,\cite{Landau2}
the quantum mechanical features of each process are non-negligible, and the 
sound attenuation is dependent on phonon energy or frequency, but not the temperature.  
Landau derived a result which covers both limits continuously,\cite{Landau2} namely
\begin{equation}
\alpha_{0}(P,T,\omega) = \alpha^{\prime}(P)\,T^{2}\,
\left \{ 1\,+\, \left ( \frac{\hbar \omega}{2 \pi k_{B}T} \right )^{2} \, \right \}
\; \textrm{for}\; k_B{T} \ll \hbar\omega \ll k_{B}T_{F}\;.
\end{equation}
Thus, the transition from the classical to the quantum regimes is expected to 
occur when 
$\hbar\omega \approx 2\pi k_{B}T$.  This result has been discussed often in the 
literature;\cite{Baym,Dobbs,Abrikosov,Wilks,Serene} 
 however, in spite of efforts by (probably) every experimental group 
that ever 
propagated ultrasound in normal liquid $^{3}$He below 10 mK, this prediction remains 
experimentally unverified, although some preliminary evidence for the ``quantum'' 
limit has been reported.\cite{Matsumoto1,Matsumoto2,Barre,Granroth}
  These preliminary reports are summarized in Table~\ref{table1}, where 
the theoretically predicted 
quantum term is calculated for comparisons of its size to unity, Eq.~(3).

The reasons for the null or quantitatively uncertain experimental 
results are easy to understand.  Ideally speaking, the highest frequencies 
($\gg 1$ GHz) and the lowest temperatures ($\ll 1$ mK) are required to conclusively 
confirm the quantum limit.  Unfortunately, the highest frequency 
measurements are limited by experimental constraints and the 
lowest temperatures are bounded by the superfluid transition.  
Consequently, measurements have been restricted to the 
\emph{``transition region''}, \emph{i.e.} between the classical and 
quantum limits, where the absolute values of attenuation are necessary 
for quantitative arguments (see Table~\ref{table1}).  
Within the \emph{``transition region''}, Eq.~(3) is considered 
to be valid, but a detailed study has not yet been made.
One way to think of the experiment is in the 
form of the measured attenuation, $\alpha_{m}$, plotted as a function of $T^{2}$.  
In other words, rewriting Eq.~(3) gives
\begin{equation}
\alpha_{m}(P,T,\omega)\,=\,\alpha^{\prime}(P)\,T^{2}\,+\,
\alpha^{\prime}(P) \, \left ( \frac{\hbar \, \omega}{2\pi \, k_{B}} \right )^{2} \,+\,
\alpha_{bk}(\omega)\;\;\;,
\end{equation}
where $\alpha_{bk}$ is the  parasitic background attenuation and all the 
temperature, pressure, and frequency dependences have been shown explicitly.  
For a positive result, the experiments must unambiguously identify a non-zero 
y-intercept, on an $\alpha_{m}$ {\emph vs.} $T^{2}$ plot, that is larger 
than the background contribution.  The major obstacle in all of the experiments 
(Table~\ref{table1}) is to reliably determine the background attenuation, and 
undoubtedly more effort is required to overcome the difficulties in determining 
the absolute values of the attenuation.  Any possible pressure 
dependence of $\alpha_{bk}$ will be discussed when appropriate.

\section{OVERVIEW OF GENERAL APPROACH}	

The general strategy for measuring the quantum term in zero sound uses the 
temperature dependence of first sound to calibrate the signal levels and involves 
several steps.  Essentially, we must fit the temperature dependence of the first 
sound amplitude to a known expression to determine the absolute attenuation in 
both the first and zero sound regimes.  The process of converting the raw zero 
sound data into absolute attenuation is summarized in the flow chart shown in 
Fig.~1.  Here, we sketch the general plan, and the details for will be 
given in the following sections.

Naturally, the first steps involve the acquisition of the experimental signals, 
and standard procedures are used to subtract crosstalk and to integrate the 
received pulses.  Further analysis requires values for the longitudinal 
viscosity at 1 bar and 5 bar, and we have used the data reported by 
Bertinat \emph{et al.}\cite{Bertinat} and Nakagawa \emph{et al.}\cite{Nakagawa}  
Using this viscosity, we subtract a 
correction due to the walls of the cylindrical sample cell from the first sound raw 
data.  Applying the wall correction at this stage of the analysis 
decreases the number of independent 
variables and improves the quality of the subsequent fits.  Next, the temperature 
dependence of the first sound data must be fit to obtain the frequency 
dependent factor, $F(\omega)$.  The fits were performed on the natural logarithm 
of the amplitude as a function of temperature so that the factor, $F(\omega)$, 
represents a constant vertical shift.  Several additional properties of the liquid 
are required to calculate the 
first sound attenuation, including density, $\rho$, thermal conductivity, $\kappa(T)$, 
isothermal and isobaric specific heats, $c_{v}(T)$ and $c_{p}(T)$, first sound 
velocity, $C_{1}$,  and second viscosity $\eta(T)$.  The first 
sound velocity, $C_{1}$, is determined directly from the data by measuring 
time-of-flight of first sound pulses.  The second viscosity is generally assumed to be small 
and negligible relative to the other contributions.\cite{Baym,Sykes}   
The remaining properties have been calculated 
by Kollar and Vollhardt\cite{Kollar} using a consistent set of thermodynamic quantities.  
The pressure dependent factor $\alpha^{\prime}(P)$ is determined by a linear 
fit to the natural logarithm of the zero sound data plotted as a function of the 
square of temperature.  Once this value and the frequency dependent factor 
$F(\omega)$ are known, the zero sound data can be dissected for evidence of additional 
attenuation that may be due to the quantum limit.

\section{EXPERIMENTAL DETAILS}

The experiment was performed using the same cylindrical sample cell employed in 
the previous experiment;\cite{Granroth} however, the transducers and 
{\textsc macor} spacer are different.  
Details of the acoustic technique, including the transducers, spectrometer, 
cell dimensions, and thermometry have been reported 
elsewhere,\cite{Masuhara2,Watson} so only the salient aspects will be given~here.  

\subsection{Sample Cell and Thermometry}
The experimental activity was divided into two separate stages with the sample 
cell pressure fixed at either 1 bar or 5 bar.  The cell pressure was changed 
using a standard zeolite 
absorption dipstick and was monitored by a Digiquartz transducer.  
After the desired pressure was set, a room temperature valve was closed at the top 
of the cryostat.  For this experiment, the pressure was not measured 
\emph{in situ}, but a low temperature pressure gauge was added and used in 
subsequent work.  The volume of the capillary line between the top of the cryostat 
and the $^{3}$He cell will produce a small change in the cell pressure which will 
be considered in our analysis.

The sample cell was placed in a silver tower mounted on a copper plate attached to the 
top of the copper demagnetization stage.  Further details of this cryostat design are 
described elsewhere.\cite{Xu}  A silver powder heat exchanger in the silver tower 
provides thermal contact for cooling the liquid sample.  Miniature coaxial 
cables with a superconducting core and a CuNi braid were used between the tower 
and the 1 K pot.  Stainless steel semi-rigid coaxial cables were used from the 1 K 
pot to the room temperature connectors.  Above 40~mK, the temperature was measured 
using a calibrated ruthenium oxide (RuO$_{2}$, Dale RC-550) resistor which has 
a resistance of $\approx 500$ $\Omega$ at ambient temperature.\cite{Meisel}  
The value of this 
RuO$_{2}$ thermometer was measured using an AC resistance bridge 
(Linear Research).  A heater mounted on the nuclear stage supplied up to 1 $\mu$W 
of thermal power.  From 40 mK to 1 mK, the temperature was measured using a 
$^{3}$He melting curve thermometer.  This experiment used the $^{3}$He melting curve 
described by Ni \emph{et al.},\cite{Ni} which is consistent with the 
Greywall scale.\cite{Greywall1}  
Since the melting curve determined by Ni and co-workers did not extend 
much above the minimum in the melting curve, the scale given by Grilly\cite{Grilly} 
was used above 325 mK.  The resultant temperature calibration 
from the Grilly scale was adjusted by a 
constant to match the value of the minimum of melting curve as given by 
Ni \emph{et al.}  Below nominally 3 mK, a Pt wire NMR thermometer (PLM-3, 
Instruments of Technology, Finland) was used and calibrated against the $^{3}$He 
melting curve thermometer. 

\subsection{Acoustic Techniques, Path Length, and Dynamic Range}
A commercial Tekmag NMR spectrometer was used for the present work, and 
this arrangement allowed for phase sensitive detection.\cite{Watson} 
Our configuration contrasts with the experiments conducted by 
Granroth \emph{et al.},\cite{Granroth} who employed Matec electronics (Models 
310 and 625) for amplitude detection.  For our study, a 
typical pulse sequence consisted of 64 pulses with 4 s delay between each pulse.  
A phase cycling procedure was used for averaging over multiples of four pulses and a 
relatively long 4~$\mu$s pulse was used to limit the frequency bandwidth of the 
transmitted pulses.  The in-phase and quadrature components were separately 
digitized at 10 Msamples/s and 2048~samples were recorded for each pulse.  
Between each pulse sequence, there was at least 8~minutes delay to avoid heating the liquid.  
Before and after the acquisition of each data set, the temperature value was 
recorded and averaged to account for any temperature drift.  The output level of 
the spectrometer was fixed at approximately 13 dBm (maximum) during the experiment 
so that comparisons between received signal levels could be performed.  To avoid 
heating the liquid, a variable inline attenuator (set at $- 20$ dB) 
was used on the transmitter side so that the cumulative attenuation from the 
cables provided a pulse input power of approximately 1 nJ.  To test for 
linearity, signals were averaged at fixed frequency and temperature using 
different attenuator values.\cite{Watson} 

The received amplitude of a 16 MHz pulse as a function of time at a pressure of 
1 bar and a temperature of 1.11 mK is shown in Fig.~2(A).  The coherent noise due 
to electrical feedthrough is evident in the first 10 $\mu$s of data.  To improve 
the signal to noise ratio, a reference signal was taken at 3.52 mK (Fig.~2(B)), 
where the received pulse amplitude cannot be differentiated from the noise, and was 
subsequently subtracted from all of the zero sound data at 1 bar.  The real and 
imaginary portions of the signal were subtracted separately before calculating the 
magnitude.  The result after subtraction is shown in Fig.~2(C), where the crosstalk 
feature at approximately 7 $\mu$s has clearly been removed.  Prior to any further 
analysis, a similar procedure was performed for the zero sound data at 5 bar and 
the first sound data at both pressures.  The integration of the received pulses was 
performed over only a 4 $\mu$s area.  Furthermore, the uncertainty for each 
integration was determined by integrating a 4 $\mu$s area located near 
$200$ $\mu$s, \emph{i.e.} near the end of the data acquisition.

The absolute attenuation is linearly dependent on the path length, and therefore 
an \emph{in~situ} measurement of this length is desirable.  The response observed 
from a 22~MHz pulse at 5~bar and 1.55~mK is shown in Fig.~3.  The time length 
between successive echoes, combined with the tabulated velocity of zero 
sound,\cite{HV} allow us to calculate a path length, $\ell$, of
\begin{equation}
\ell = 0.327 \pm 0.002 \, \mathrm{cm} \,\,\,.
\end{equation}
The uncertainty in this measurement is primarily determined by the 100 ns time 
resolution of the spectrometer.  This determination of $\ell$ was then combined 
with the data in the first sound regime to obtain values for $C_{1}(P,T)$.  
These values are in agreement with published values, within the uncertainty.  
The temperature 
and pressure dependences of $C_{1}$ are required for the calibration of the absolute 
attenuation, and additional details are given in Appendix~B.

The dynamic range of our attenuation measurement is restricted by our input power 
and cell size.  
Figure 4 shows the attenuation as a function of the logarithm of temperature at a 
pressure of 1 bar and a frequency of 16 and 64 MHz.  The attenuation on the left 
and right side of Fig.~4 is due to zero sound and first sound, respectively.  The 
absolute attenuation (y-axis) has been determined by fitting the temperature 
dependences of the zero sound and first sound data to known expressions.  If the 
attenuation is smaller than approximately 0.4 cm$^{-1}$, then changes in the 
attenuation will be smaller than the scatter due to the noise (lower horizontal 
line).  If the attenuation is larger than 16.2 cm$^{-1}$ then the signal will not be 
detected (upper horizontal line).

\section{ZERO SOUND}

Typical received amplitudes as a function of time and temperature for  zero 
sound at 10~MHz and 1 bar are shown in Fig.~5.  The crosstalk has been 
subtracted using the method discussed in the previous section.  The decrease in 
amplitude is due to the $T^{2}$ dependence of attenuation on temperature.  At low 
frequency, 8, 10, and 16 MHz, the transducer response is broader and hence the 
received pulses resemble the square transmitted pulse.  Higher frequency received 
pulses, $f > 19$ MHz, contain sharp peaks related to the transducer responses at 
that frequency.\cite{Masuhara2}

In order to accurately measure the absolute attenuation of zero sound, we must 
determine the large frequency dependence due to the transducer response.  
Consequently, it is necessary to compare the zero sound signal levels with the 
high temperature, low attenuation limit of first sound.

\section{FIRST SOUND}

In order to accurately determine the zero sound attenuation, we must calibrate 
the large frequency dependence of the transducer response by studying the attenuation 
of first sound.  Herein lies the greatest uncertainty or propagation of systematic errors.  

The first sound attenuation was measured at fixed temperatures in the range 
from approximately 30~mK to 800~mK at two different pressures.
For these measurements, 21 and 27 
discrete temperatures were used at 1 bar and 5 bar, respectively. As in the previous section, low 
temperature $(T \approx 30$~mK), high attenuation data were subtracted from all other 
data to eliminate the contribution from electrical crosstalk.  
The integration of the received signals was performed over a 4~$\mu$s window which 
was adjusted to compensate for the pressure being studied since the speed of sound changed.  
The noise level for each integration was produced by a further integration over a 
4~$\mu$s region at $t = 200$~ $\mu$s.

The attenuation of first (hydrodynamic) sound can be written as
\begin{equation}
\alpha_1(P,T,\omega) = 
\frac{\omega^{2}}{2\,\rho \,C^{3}_{1}} \left[ \left(\frac{4}{3}\eta + \zeta \right)
\;+\;\kappa\left(\frac{1}{c_{v}}-\frac{1}{c_{p}}\right) \right]\;\approx\; 
\frac{2\,\omega^{2}\, \eta}{3 \,\rho \,C^{3}_{1}}\;\propto\;\frac{\omega^{2}}{T^{2}}
\;\;\;,
\end{equation}
where $\eta$ and $\zeta$ are the first (longitudinal) viscosity and second (bulk) 
viscosity, $\kappa$ is the thermal conductivity, $\rho$ is the density, $C_{1}$ 
is the sound velocity, and $c_{v}$ and $c_{p}$ are the specific heats at constant 
volume and pressure.\cite{LLbook}  As mentioned previously, the first sound velocity, $C_{1}$, 
is determined using the time delay between the arrival of the received pulse and 
the length of the cell.  The thermal conductivity, $\kappa$, and the specific heat at 
constant volume, $c_{v}$, have been measured by Greywall.\cite{Greywall1,Greywall3}  
Using the measurements 
of Greywall, Kollar and Vollhardt\cite{Kollar} have calculated the density, $\rho$, 
and specific heat at constant pressure, $c_{p}$, using a consistent set of 
thermodynamic quantities.  The contribution to the attenuation from the thermal 
conductivity is expected to be significant as it has a linear dependence on 
temperature and a quadratic dependence on frequency.  Therefore, a term 
corresponding to the thermal conductivity must be included when fitting 
the temperature dependence of first sound attenuation to a known expression.

An additional attenuation correction, which accounts for scattering of the 
quasiparticles on the walls of the cylindrical cell,\cite{LLbook,Eska} can be written as
\begin{equation}
\alpha_{wall}=\frac{\omega}{2\, R \, C_{1}} \left[\frac{2\,\eta}{\rho \, \omega} \right]^{1/2}
\;\;\;,
\end{equation}
where $R$ is the radius of the cell.  For a frequency of 10 MHz and a temperature of 
800~mK, the attenuation due to the wall scattering is 14\% of the hydrodynamic 
attenuation.  To improve fitting accuracy, for all of the data presented, the wall 
attenuation was subtracted from the data before fitting to the attenuation in 
Eq.~(6).  
By combining Eq.~(6) with an additional frequency dependent factor, we can write 
the amplitude of first sound, ${\mathcal A}_1$, (recalling that the crosstalk and wall attenuation 
corrections have already been made)  as
\begin{equation}
{\mathcal A}_{1}\,=\,
F(\omega)\, \exp \left\{ -\,\frac{\omega^{2} \, \ell}{2\,\rho \,C^{3}_{1}} 
\left[ \left(\frac{4}{3}\eta + \zeta \right)
\;+\;\kappa\left(\frac{1}{c_{v}}-\frac{1}{c_{p}}\right) \right] \right\} \;\;\; ,
\end{equation}
where $F(\omega)$ is the frequency dependence of the transducer and the related 
electronics.  The factor $F(\omega)$ is the only unknown quantity in Eq.~(8) and 
hence is uniquely determined by fitting Eq.~(8) to the data.  It is important to 
note that the terms inside the exponential in Eq.~(8) decrease as the temperature 
is increased but they do not become arbitrarily small.  
In other words, $F(\omega)$ cannot be determined by simply taking the 
$T \rightarrow 0$  limit.  
Finally, once $F(\omega)$ is determined from Eq.~(8), the amplitude of zero sound 
can be written as 
\begin{equation}
{\mathcal A}_{0}\,=\, 
F(\omega)\, \exp \left\{ -\, \alpha^{\prime}(P)\, T^{2} \, \ell \, 
\left[ 1\,+ \,\left( \frac{\hbar \omega}{2\pi k_B T} \right)^2 \right] \right\} \;\;\;.
\end{equation}

The longitudinal viscosity, $\eta$, has been measured by 
several research groups\cite{Bertinat,Nakagawa,Black,Carless} 
over various temperature ranges and pressures.  At temperatures above 100 mK, 
experimental data\cite{Bertinat,Black} indicate that $\eta$ deviates 
from pure $1/T^2$ behavior, and the temperature dependence of $\eta$ has the 
form
\begin{equation}
\eta(P)\,=\,\frac{\Gamma_{1}(P)}{T^{2}}\,+\,\frac{\Gamma_{2}(P)}{T^{n}} \;\;\;,
\end{equation}
where $\Gamma_{1}$, $\Gamma_{2}$, and $n$ are constants.  
We are unable to determine 
the second constant, $\Gamma_{2}$, directly from our data, 
and both previous measurements\cite{Bertinat,Black} of this 
term have been performed only at saturated vapor pressure.  
Consequently, some assumptions and an interpolation of the reported 
viscosity values are required.
   
For our analysis, the first constant, $\Gamma_1(P)$, 
was taken from the results of 
Nakagawa \emph{et~al.}\cite{Nakagawa} who worked over a wide range of pressures.  
The second constant, $\Gamma_2$, was obtained from the value reported by 
from Bertinat \emph{et al.},\cite{Bertinat} after it was multiplied by the 
same factor necessary to make their $\Gamma_1$ term equal to the $\Gamma_1$ 
term reported by Nakagawa \emph{et al.} at either 1 bar or 5 bar.  
This correction was necessary since Bertinat \emph{et al.} only worked at 
saturated vapor pressure.  
With this approach, we are assuming that the pressure dependences of $\Gamma_1$ 
and $\Gamma_2$ are identical and that the temperature dependence of 
the second term, $1/(T^{0.42})$, is independent of pressure.
With this interpolation procedure, the final result for the viscosity 
expressed in poise (P) is 
\begin{equation}
\eta(P=1\, \textrm{bar})\,=\,\frac{1.99}{T^{2}}\,+\,\frac{4.44 \times 10^{-4}}
{T^{0.42}}
\end{equation}
at 1 bar and 
\begin{equation}
\eta(P=5\, \textrm{bar})\,=\,\frac{1.62}{T^{2}}\,+\,\frac{3.62 
\times 10^{-4}}{T^{0.42}}
\end{equation}
at 5 bar, where $T$ has units of mK.  We have more confidence in the value of 
$\Gamma_{2}$ at 1 bar than at 5 bar because the pressure is closer to 
saturated vapor pressure. 

An expression for the second or bulk viscosity, $\zeta$, has been derived by 
Sykes and 
Brooker\cite{Sykes} in terms of a collision integral; however in their analysis, no 
attempt was 
made to evaluate the expression numerically.  Although it has generally been 
assumed that 
the second viscosity has a $\zeta \sim T^{0}$ dependence, they predict that 
the second viscosity has a 
$\zeta \sim T^{2}$ dependence.  It should be noted that their calculation 
relies on the assumption that 
$^{3}$He is well described by Fermi liquid theory which is not necessarily 
valid above the 
temperature of approximately 150 mK $(T \approx T_{F}/10)$.  In addition,  
the magnitude of the 
second viscosity at temperatures below 1 K is expected to be 
negligible,\cite{Wilks} so we have not included it in our analysis.

To summarize, it is noteworthy that Eqs.~(11) and (12) are 
self-consistent with our measurements of the attenuation of first 
sound.\cite{Watson}  This point is illustrated in Fig.~6, where the first sound 
data are well described by Eq.~(8) when using the viscosity given by 
Eq.~(11).  Alternatively, if the values for the viscosity are 
interpolated from the data reported by Bertinat~\emph{et al.}\cite{Bertinat} 
and Carless~\emph{et al.},\cite{Carless} as was done by Granroth and 
co-workers,\cite{Granroth} the present data are not well fit, Fig.~6.  
The present work represents an improvement over the results reported by 
Granroth~\emph{et al.} since the number of temperatures investigated was 
increased by approximately 3 and the uncertainty limits were reduced 
by a factor of about 2.

\section{ADDITIONAL UNCERTAINTIES AND ANALYSIS}

To this point, our analysis have not addressed three important sources of 
uncertainty, namely the nonparallelism of the transducers, the measurement and 
stability of the pressure, and the variation of the acoustic impedance.  

The nonparallelism of the 
transducers will add 
an additional frequency dependent factor, $N(\omega)$, to the zero and first sound 
amplitudes, so the received amplitude is related to the transmitted 
amplitude as\cite{Abraham,Truell}
\begin{equation}
{\mathcal A}_{rec}\,=\, {\mathcal A}_{trans}\,N(\omega)\, 
e^{-\, \alpha\, \ell} \;\;\;.
\end{equation}
The factor $N(\omega)$ is expressed in terms of a Bessel function
\begin{equation}
N(\omega)\,=\,2\,\frac{J_{1}\,\left[(2m-1)\,\frac{\omega}{C}\,R\,\Theta \right]}
{(2m-1)\,
\frac{\omega}{C}\,R\,\Theta} \;\;\;,
\end{equation}
where $\Theta$ is the angular error and $m$ is the received pulse number.  
In a previous experiment,\cite{Granroth} 
the nonparallelism was determined by filling the cell with $^4$He and 
measuring the 
attenuation of closely spaced frequencies from 8 to 64 MHz.  However, for 
the present experiment, 
the transducer properties could not accommodate a continuous 
sweep of the 
frequency.  Therefore the uncertainty due to nonparallelism must be estimated.  
Fortunately, the uncertainty 
due to nonparallelism will occur in both the zero and first sound regimes, 
and the 
effect will partially cancel.  Nevertheless, the sound velocity change 
between zero and first 
sound will cause the zeros in the Bessel function to appear at slightly 
different frequencies.  
A relatively small increase in the nonparallelism will result in a large 
increase in the uncertainty at 
high frequencies.  In addition, at specific frequencies that correspond 
to a zero in the 
Bessel function, the uncertainty will increase dramatically.  The uncertainty 
at 5 bar is less than the uncertainty  
at 1 bar because there is less difference between the first sound 
and zero sound 
velocities at that pressure.  In the previous experiment by 
Granroth \emph{et al.},\cite{Granroth} the value of 
$\Theta$ was measured as $(4.1 \pm 0.1) \times  10^{-4}$ radians, which is 
consistent with uncertainties due to 
machining.  For the present experiment, the nonparallelism is assumed to be the 
same as the 
previous measurement by Granroth \emph{et al.} and the corresponding 
uncertainty is estimated as 
$\pm$ 20\% of the final result at 1 bar and $\pm$ 10\% at 5 bar.  These 
percentage values appear to be large only 
because the quantum term in zero sound attenuation is exceedingly small.

There are two sources of uncertainties that are related to pressure changes in 
the cell.  
First, there are pressure changes due to variations in temperature.  
Second, there are 
pressure changes due to the cyclical change (period of 3 days) of the 
$^{4}$He level in the 
dewar.  The pressure change produced by a variation in temperature was 
measured in a 
subsequent experiment using a strain gauge attached to the $^{3}$He cell.  
It is difficult to 
calculate the effect of the pressure change on the transducer properties.  
However, we can 
estimate the uncertainty in the amplitude due to the temperature variation as 
approximately $\pm$ 4\% in the first sound regime.  In comparison, the pressure 
change due to the $^{4}$He 
level in the dewar is significantly less, with the error of approximately 
$\pm$ 1\% of the first 
sound amplitude.

There is an additional correction due to a change in the acoustic impedance of 
liquid $^{3}$He.  The amount of energy that is transmitted into the $^{3}$He 
cell is a function of the 
impedance of the transducers and the $^{3}$He liquid.  The real component of 
the impedance of 
a hydrodynamic fluid can be written as
\begin{equation}
\textrm{Re}[Z]\,=\,\rho(T,P) \, C(T,P) \;\;\;,				
\end{equation}
where $\rho(T,P)$ is the density of the liquid and $C(T,P)$ is the sound velocity.  
Between zero 
sound and first sound, there is a significant increase in the sound velocity 
and a subtle 
change in $\rho$ combining to cause an increase in acoustic impedance.  
Independent of 
frequency, the overall signal amplitude will decrease in the first sound regime.  
Accordingly, a constant adjustment must be applied to the first sound data 
to force a zero 
intercept in a linear fit of the data in Figs. 7 and 8.  The constant 
adjustment is 0.466 
cm$^{-1}$ and 0.537 cm$^{-1}$ for the 1 bar and 5 bar data, respectively.  
The ratio between these two 
values (0.537/0.436) = 1.15 is roughly equal to the ratio between the 
calculated values of 
[$\rho \, C$(5 bar)]/[$\rho \, C$(1 bar)] = 1.38.

\section{FINAL RESULTS AND DISCUSSION}

After all the analysis, the received zero sound 
signals yield a corrected attenuation, $\alpha_{\mathrm{cor}}$, which 
should represent the absolute attenuation of the liquid.  From Eq.~(3), 
the final results may be plotted in the form  
${[\alpha_{\mathrm{cor}}/\alpha^{\prime}\,T^{2}]\,T^{2}-1}$  {\emph vs.} 
$(\omega/2\pi)^{2}$, 
as shown in Fig.~7 ($P = 1$ bar) and Fig.~8 ($P = 5$ bar).  According to Landau's 
prediction, the data are expected to fall on a straight line with a slope of 
$(\hbar/k_{B})^{2} = 5.83 \times 10^{-5}$ (mK/MHz)$^{2}$.  With a nonlinear 
least-squares fit of the 1 bar results, where each data point is 
weighted by the inverse square of its 
uncertainty, the slope of the fitting line is $(1.3 \pm 0.3)(\hbar/k_{B})^{2}$, and 
this result is consistent with Landau's prediction.
The error limits that are given encompass all of the theoretical 
and experimental 
uncertainties including the non-parallelism, the pressure, the 
value of $\alpha^{\prime}(P)$, 
and the path length, $\ell$.  
An unconstrained fit of the 5 bar data gives an unphysical negative slope, which 
means that we have  
not taking into account a source of attenuation in the first sound regime at this 
pressure.  Constraining the fit to omit non-negatitive results yields a slope of 
$(0.0 \pm 1.4)(\hbar/k_{B})^{2}$ at 5 bar.  
The quantity that is most uncertain at 5 bar is the second term in 
the longitudinal 
viscosity, $\Gamma_{2}$ (see Eq.~(12)), and our assumptions about its pressure 
dependence may be responsible for the large uncertainty at 5 bar.

If the viscosity is taken by an interporlation of the data of 
Bertinat~\emph{et al.}\cite{Bertinat} and Carless~\emph{et al.}\cite{Carless} 
as was done by Granroth~\emph{et al.},\cite{Granroth} 
rather than from Eq.~(11), then the present data would yield a 
slope of $(6.2 \pm 0.6)(\hbar/k_{B})^{2}$ at 1 bar.  This value is consistent with 
the $(5.6 \pm 1.2)(\hbar/k_{B})^{2}$ result 
reported by Granroth \emph{et al.}  However, it is important to stress that 
both of these values arise from analysis that 
employs a form of the viscosity that provides a 
poor description of the attenuation of first sound, as discussed at the end of 
Section VI.  Consequently, the present work is an improvement over the older results 
due to number of reasons, including a more thorough study of the attenuation of 
first sound. 

\section{CONCLUSIONS}

In conclusion, to within one deviation of the overall uncertainty 
of the measurements, the results at 1 bar agree with the predicted 
quantum term in zero sound predicted by Landau.  
The corresponding data and analysis at 5 bar yield relatively large uncertainties 
and consequently provide a null result for the measurement of the quantum term.  
The overall uncertainty at 5 bar arises mainly from the lack of a detailed 
knowledge of the second term in the longitudinal viscosity.  
Nevertheless, it is noteworthy that a  similar result was obtained by 
Matsumoto \emph{et al.},\cite{Matsumoto2} who 
measured the predicted quantum term at 1 bar and a null result at 5 bar.  
Improvements in measuring the absolute attenuation over expanded frequency and  
pressure ranges are required to 
make further progress on this fundamental issue.

\appendix
\section{QUASIPARTICLE SCATTERING AMPLITUDE}

In the zero sound regime and after subtracting the electrical crosstalk, 
the received signal was integrated inside the 4 $\mu$s window corresponding 
to the received pulse, Fig.~2.  
Neglecting the possible 
attenuation coming from the quantum limit, the integrated amplitude, 
${\cal A}$,  may be written as
\begin{equation}
{\cal A}\,=\,\beta\,\exp(- \alpha^{\prime}(P)\,T^{2}\,\ell)\;\;\;,
\end{equation}
where $\alpha^{\prime}(P)$ represents the contribution from quasiparticle 
scattering, Eq.~(1). 
In Fig.~9, the natural logarithm of the integrated amplitude at 
1 bar is plotted as a function of $T^2$, and the value of 
$\alpha^{\prime}(P)\,\ell$ is 
determined by a linear fit of this data.
Since the value of $\ell$ is established 
\emph{in situ} by time-of-flight analysis (Eq.~(5)), a value for $\alpha^{\prime}(P)$ 
may be obtained.  For most frequencies, the measured values of $\alpha^{\prime}(P)$ 
are equal to within the experimental uncertainty, however there are 
slight variations due to subtle pressure changes.  Similar data were obtained 
at 5 bar (Fig.~10), and all of the data are tabulated in Table~\ref{table2}.  

The value of $\alpha^{\prime}(P)$ can be compared to values obtained  
from previous experiments,\cite{Abel2,Lawson,Ketterson,Mast,Nara,Avenel} 
and Fig.~11 shows the reported values of $\alpha^{\prime}(P)$.  
An alternative and perhaps physically more meaningful presentation of
$\alpha^{\prime}$ is as a function of molar volume, $\nu$.  Using the 
tabulated $P(\nu)$ function,\cite{HV} $\alpha^{\prime}(P)$ can be 
converted to $\alpha^{\prime}(\nu)$, and the results are  
shown in Fig.~12, where the solid 
line represents the results of second order polynomial fit of the data, 
\emph{i.e.}
\begin{equation}
\alpha^{\prime}(\nu) = 1.4088 - 0.14882 \,\nu + 4.1303 \times 10^{-3} \,\nu^{2}\;\;\; .
\end{equation}
This equation can be converted back to a pressure dependence using the known 
$\nu(P)$ relationship,\cite{HV} yielding the solid line shown in Fig.~11,
\emph{i.e.}  
\begin{equation}
\alpha^{\prime}(P) = 1.4997 -0.15200 \, P+0.010058\,P^{2}-3.397 \times 10^{4}\,P^{3} 
+4.3695 \times 10^{6}\,P^{4} \,.
\end{equation}
As far as we know, this appendix presents the first comprehensive analysis of 
known values of $\alpha^{\prime}$.

\section{PATH LENGTH AND FIRST SOUND VELOCITY}

Knowing the path length and the time delay before each received 
pulse, we can calculate the velocity as a function of temperature as shown in 
Fig.~13 for the pressure of 1 bar and 5 bar.  The time delay was determined by the 
sudden increase of the sound amplitude corresponding to the arrival of the pulse 
in the 23 MHz data.  The 23 MHz data was chosen because the start of the pulse is 
easy to identify.  Received pulses with frequencies above approximately 60 MHz 
appear rounded because the higher frequency harmonics of the pulse are attenuated.  
Within the uncertainty, the velocity is constant at both pressures.  The error bars 
have been determined by a combination of the uncertainty in the path length and 
the uncertainty in the time of arrival, which is essentially one-half the time 
resolution of the spectrometer.  In this experiment, a room temperature valve was 
used to access the sample space and consequently, the pressure was slightly 
temperature dependent due to the small volume of the $^{3}$He capillary.  
At 1 bar and 
5 bar, the pressure changes from 100~to~800~mK are approximately 0.1 bar and 
0.2 bar, respectively.  This pressure change was not measured directly during this 
experiment, 
but it was observed in subsequent runs that incorporated a low temperature 
pressure gauge.  
However, because the volume of the $^{3}$He capillary is small compared 
to the volume of the cell, we can assume that the cell volume will remain constant.  
Setting of the $^{3}$He pressure was performed at the temperature of 250~mK.  
The molar volumes corresponding to 1 bar and 5 bar at 250 mK, determined using the results 
of  Kollar and Vollhardt,\cite{Kollar} are 35.6516 and 32.5883 cm$^{3}$, respectively.  Using 
these constant molar volumes, the temperature dependence of all other thermodynamic 
quantities can then be calculated.\cite{Kollar}  In Fig.~13, 
the triangles are calculated 
from the first sound velocity expression given by 
Roach \emph{et al.}\cite{Roach} using the 
adiabatic compressibility and density values from Kollar and Vollhardt.\cite{Kollar}  
These theoretical curves have been adjusted by a constant ($-7.05$~m/s for 1~bar and 
$-0.35$~m/s for 5~bar) to match the experimental velocity value at 50~mK.  
The solid lines represent the theoretical first sound velocity calculated using 
the expression given by Abraham \emph{et al.}\cite{Abraham2}  
Likewise, these theoretical 
curves have been adjusted by a constant (1.85 m/s for 1 bar and $-0.32$ m/s for 
5 bar) to match the experimental velocity value at 50 mK.  The first sound 
velocity at 250 mK given by Halperin and Varoquaux\cite{HV} is also shown (squares).  
Our fits involving the first sound velocity used the constants 200.61 m/s 
for 1 bar and 249.62 m/s for 5 bar.

\begin{acknowledgments}
During the course of this work, we have benefited from conversatation or 
correspondence with 
D. Einzel, A. Feher, G.E. Granroth, W.P. Halperin, G.G. Ihas, Y. Lee, Y. Okuda, 
J. Parpia, J.A. Sauls, 
and the late J.R. Hook.  We gratefully acknowledge N. Bushong for 
assistance in preparing Fig.~1.
\end{acknowledgments}



\begin{table}
\begin{center}
\begin{tabular}{||c|c|c|c|c||} \hline\hline
 & & & & \\
Reference	&	$\omega /2 \pi$ 	&	$T_{min}$	&	$P$ &	
$\left( \begin{array}{c} \hbar \, \omega \\ \overline{2 \pi \,k_{B}\, T_{min}} 
\end{array} \right )^{2}$\\ 
          & (MHz)         &  (mK)     & (bar) &  \\ \hline\hline
Matsumoto \emph{et al.}\cite{Matsumoto1} & 389 & 2.5 & 3  &   0.11 \\ \hline
  & 389 & 6   & 0.4&   0.25 \\ \cline{2-5} 
Matsumoto \emph{et al.}\cite{Matsumoto2} & 389 & 6   & 3  &   0.25 \\ \cline{2-5}
  & 389 & 7   & 5  &   0.18 \\ \hline
  &  84 & 7   & SVP&   0.01 \\ \cline{2-5}
Barre \emph{et al.}\cite{Barre} & 254 & 7 & SVP & 0.08 \\ \cline{2-5}
  &  422 & 7  & SVP&   0.21 \\ \cline{2-5}
  &  592 & 7  & SVP&   0.42 \\ \hline
Granroth \emph{et al.}\cite{Granroth} & 46 & 1.08 & 1 & 0.11 \\ \hline
  & 23   &   1.11   &1   & 0.03  \\ \cline{2-5}
  & 67   &   1.11   &1   & 0.21   \\ \cline{2-5}
This Work &   107   &1.11   &1&   0.54  \\ \cline{2-5}
  & 28   &   1.55   &5   & 0.02  \\ \cline{2-5}
  & 68   &   1.55   &5   & 0.11  \\  \cline{2-5}
  & 108  &   1.55   &5   & 0.28  \\ \hline\hline
\end{tabular}
\end{center}
\caption{Summary of studies made at high frequency and low temperature.  
Listed for each experiment are the frequency, the minimum temperature, 
the pressure, and the relative value of the predicted quantum term in 
the zero sound attenuation (Eq.~(3)).  The notation ``SVP'' means ``saturated 
vapor pressure''.}
\label{table1}
\end{table}

\begin{table}[p]
\begin{center}
\begin{tabular}{||c|c|c||} \hline\hline
 & &  \\
 $(\omega/2\pi)$	& $\alpha^{\prime}(P = 1$ bar)	& $\alpha^{\prime}(P = 5$ bar) \\
(MHz)	&	(cm$^{-1}$)	&	(cm$^{-1}$) \\ 
 & &  \\
 \hline\hline
8	&	&	0.91 $\pm$ 0.05 \\ \hline
10	& 1.25 $\pm$ 0.04	& 0.80 $\pm$ 0.04 \\ \hline
16	& 1.31 $\pm$ 0.02	& 0.86 $\pm$ 0.04 \\ \hline
19	& 1.30 $\pm$ 0.04	& 		  \\ \hline
20	& 1.33 $\pm$ 0.03	& 0.93 $\pm$ 0.05 \\ \hline
21	& 1.30 $\pm$ 0.04	& 		  \\ \hline
22	& 1.34 $\pm$ 0.02	&		  \\ \hline
23	& 1.32 $\pm$ 0.01	& 0.83 $\pm$ 0.04 \\ \hline
28	&			& 0.76 $\pm$ 0.04 \\ \hline
63	& 1.34 $\pm$ 0.03	& 0.94 $\pm$ 0.05 \\ \hline
64	& 1.34 $\pm$ 0.01	& 0.84 $\pm$ 0.03 \\ \hline
65	& 1.31 $\pm$ 0.04	& 0.94 $\pm$ 0.05 \\ \hline
66	& 1.28 $\pm$ 0.02	& 		  \\ \hline
66.6	& 1.35 $\pm$ 0.02	& 0.86 $\pm$ 0.02 \\ \hline
68	& 			& 0.89 $\pm$ 0.07 \\ \hline
107	& 1.4 $\pm$ 0.1		& 0.9 $\pm$  0.1 \\ \hline
108	& 1.25 $\pm$ 0.05	& 0.99 $\pm$ 0.02 \\ \hline\hline
 & &  \\
Average & 1.32 $\pm$ 0.01	& 0.88 $\pm$ 0.02 \\ 
 & &  \\ \hline\hline
\end{tabular}
\end{center}
\caption{The values of $\alpha^{\prime}(P)$, Eqs.~(1) and (16), 
as determined from a 
linear fit to the data when plotted in the manner shown in Figs. 9 and 10.}
\label{table2}
\end{table}  

\begin{figure}[p]
\centerline{\includegraphics[height=6.0in]{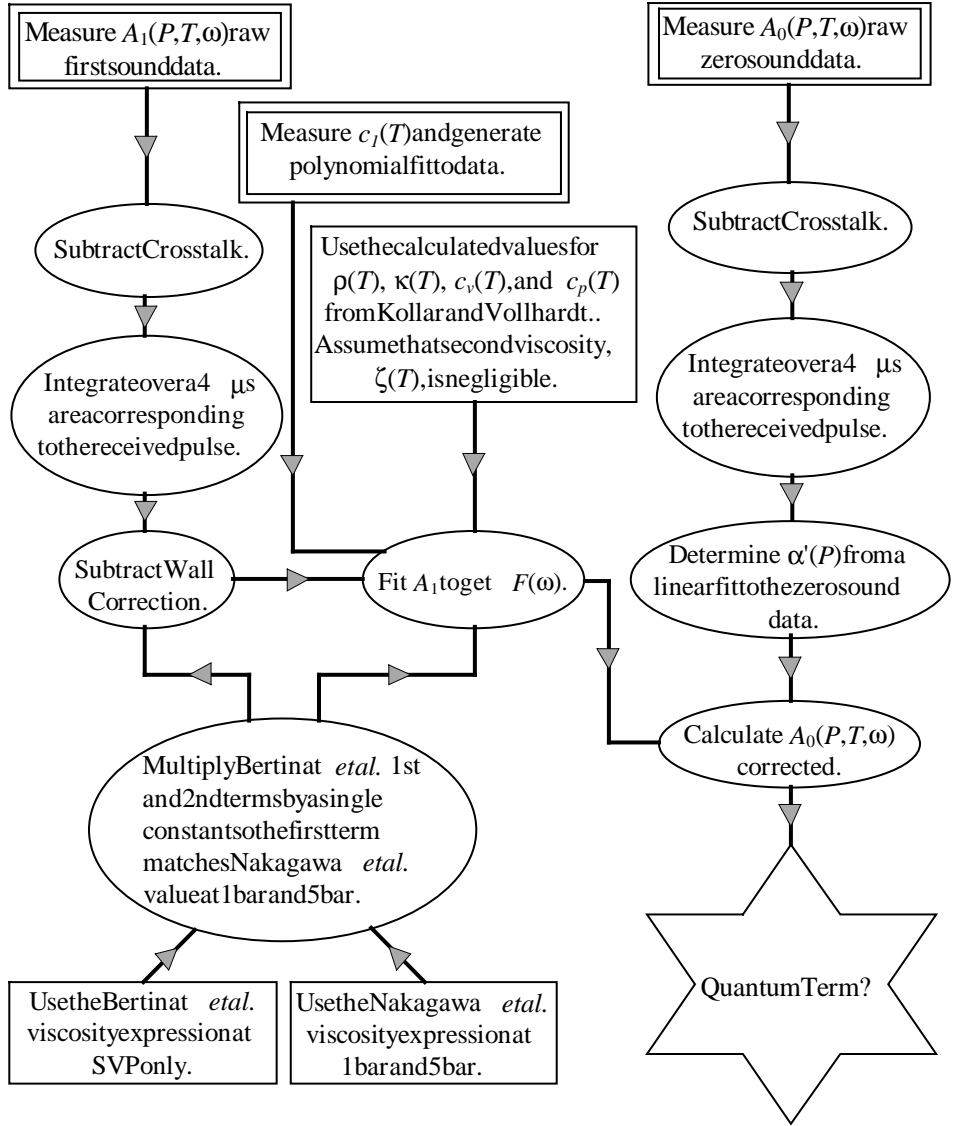}}
\caption{Flow diagram depicting the process used to analyze the data.  Input data 
from our measurements are represented by double rectangles, and 
mathematical operations are shown by ellipses.  Input about the 
properties of the liquid are bounded by rectangles, and this information was 
taken from Bertinat \emph{et al.},\cite{Bertinat} 
Nakagawa \emph{et al.},\cite{Nakagawa} and 
Kollar and Vollhardt.\cite{Kollar}}  
\label{fig:1}
\end{figure} 

\begin{figure}[p]
\centerline{\includegraphics[height=6.0in]{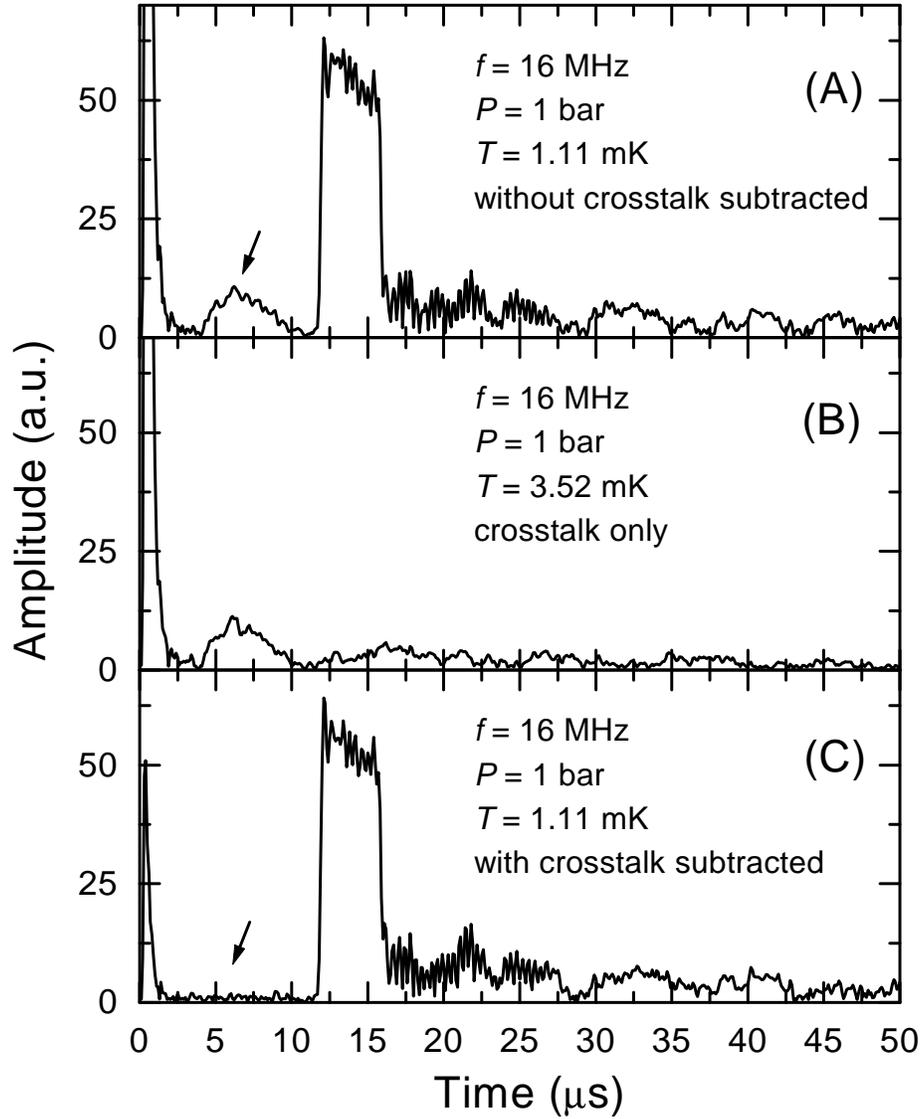}}
\caption{The received amplitude of a 16 MHz pulse as a function of 
time at a pressure of 1 bar and a temperature of 1.11 mK.  In (A),   
the coherent noise due to electrical feedthrough is evident in the first 
10 $\mu$s of data.  At a temperature of 3.52 mK (B), the signal is 
attenuated and only the crosstalk remains.  
In (C), the result of subtracting (B) from (A) is shown.}  
\label{fig:2}
\end{figure}

\begin{figure}
\centerline{\includegraphics[width=4.0in, height=3.0in]{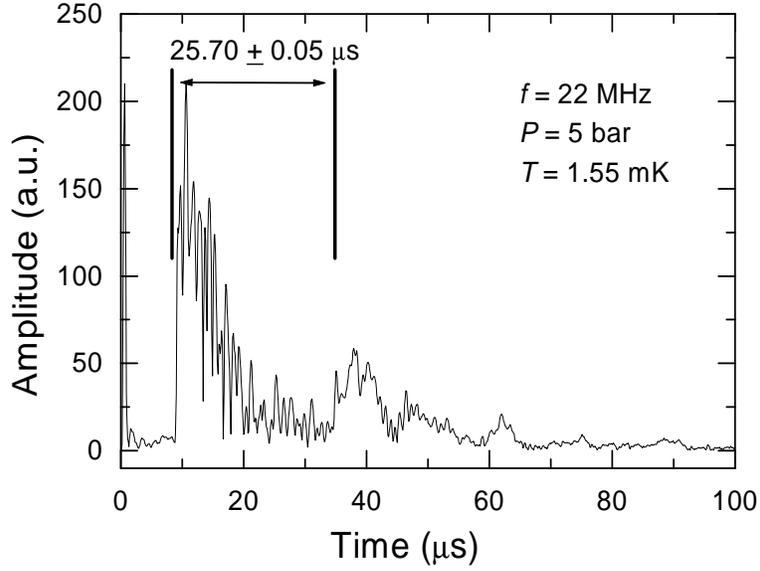}}
\caption{A 22 MHz received pulse, at 5 bar and 1.55 mK, as a function of time.  
The time length between the first received pulse and the first echo is shown.}  
\label{fig:3}
\end{figure}

\begin{figure}
\centerline{\includegraphics[width=4.0in,height=3.0in]{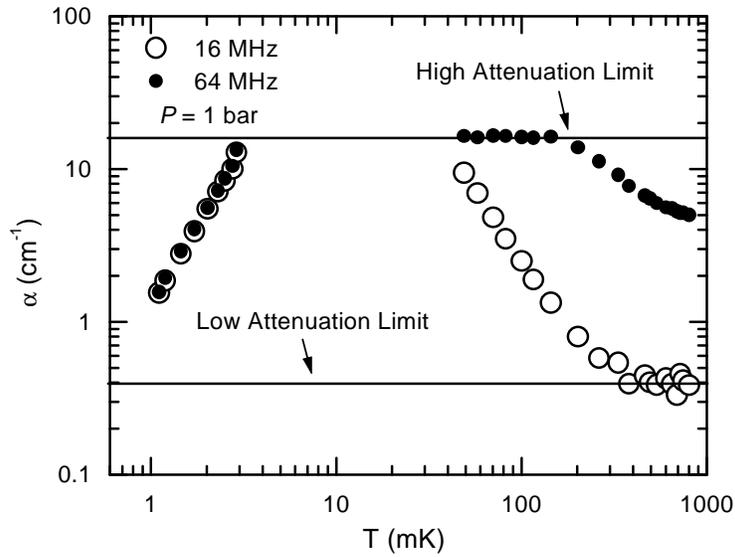}}
\caption{The dynamic range of our measurement of the 
sound attenuation as a function of the temperature at 1 bar.  
The upper horizontal line indicates the high attenuation limit where 
the signal is too small to be detected.  The lower horizontal line 
indicates the signal level where changes in the attenuation are 
smaller than the noise.}  
\label{fig:4}
\end{figure}

\begin{figure}
\centerline{\includegraphics[width=4.0in,height=3.0in]{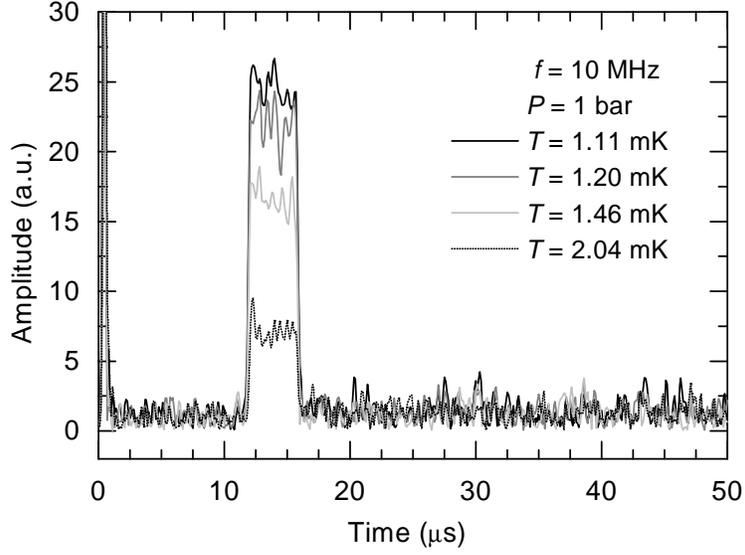}}
\caption{Typical received pulse amplitudes as a function of time and temperature 
for a zero sound pulse at 10 MHz and 1 bar.  The crosstalk has been subtracted.}  
\label{fig:5}
\end{figure}

\begin{figure}
\centerline{\includegraphics[width=4.0in, height=3.0in]{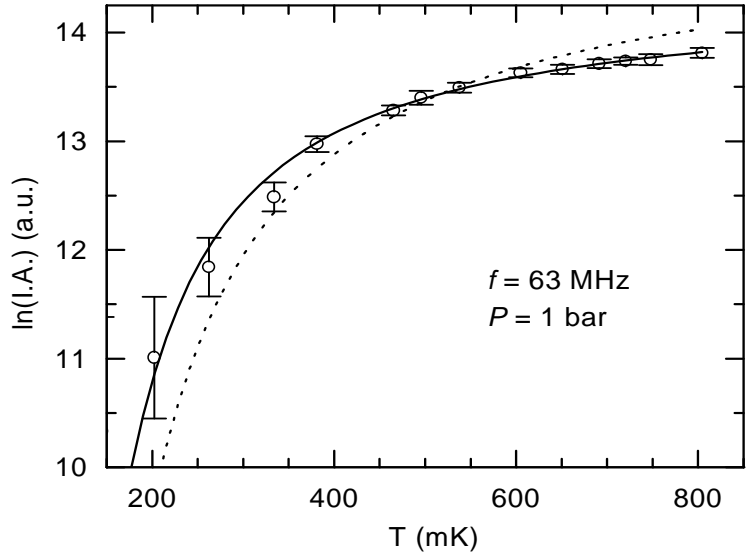}}
\caption{The natural logrithum of the integrated amplitude, $\ln$(I.A.) in 
arbitrary units, is plotted as a function of temperature at 63 MHz and 1 bar.  
The solid line is generated when using the viscosity given by Eq.~(11).  The 
dotted line arises when using viscosity values interpolated from other 
work,\cite{Bertinat,Carless} and this approach was used by Granroth 
\emph{et al.}\cite{Granroth}}  
\label{fig:6}
\end{figure}

\begin{figure}
\centerline{\includegraphics[width=4.0in, height=3.0in]{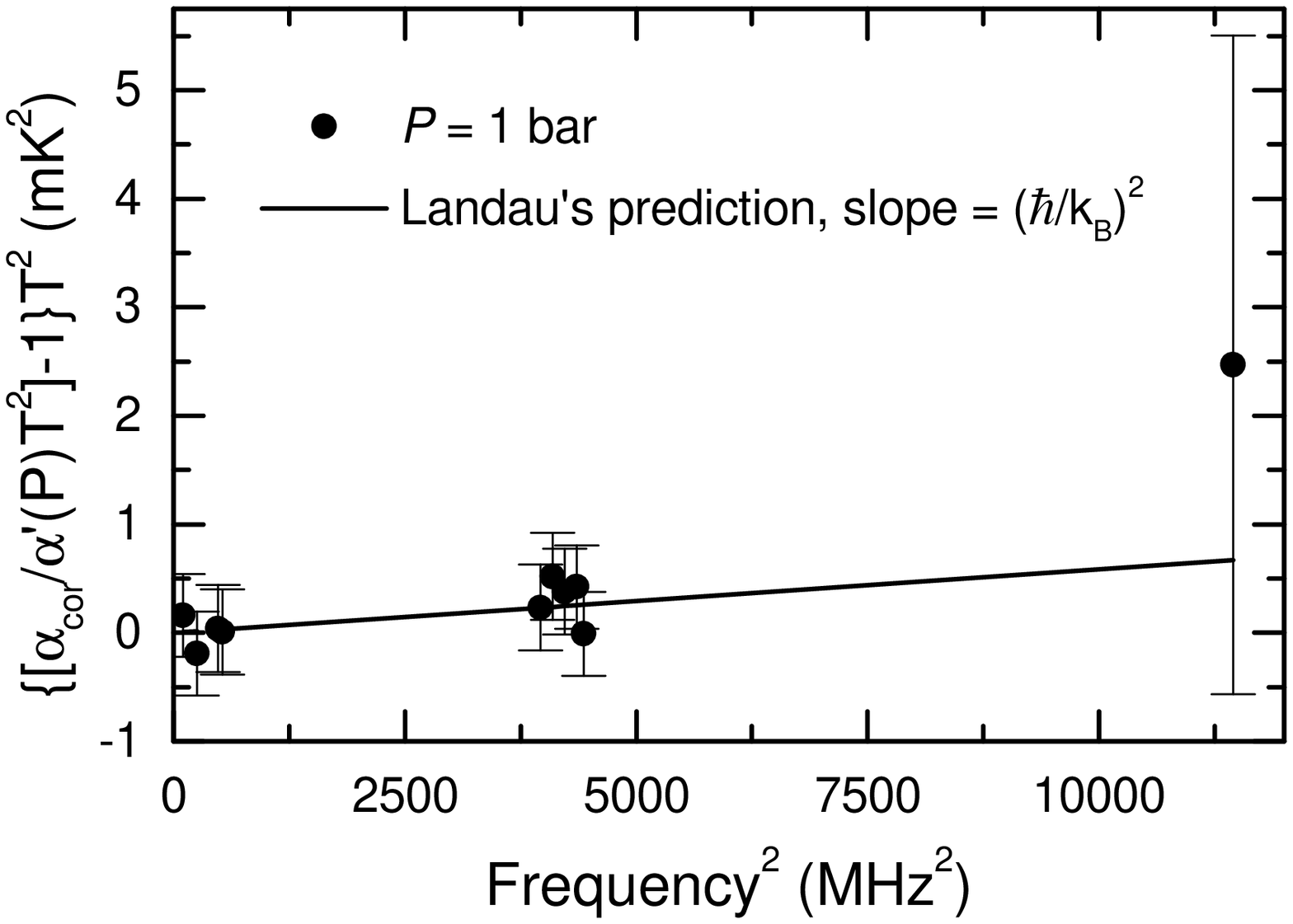}}
\caption{The corrected and normalized attenuation at 1 bar plotted 
as a function of the square of the frequency, Eq.~(1).  The data 
yield a slope of $(1.3 \pm 0.3)(\hbar/k_{B})^{2}$.}  
\label{fig:7}
\end{figure}

\begin{figure}
\centerline{\includegraphics[width=4.0in,height=3.0in]{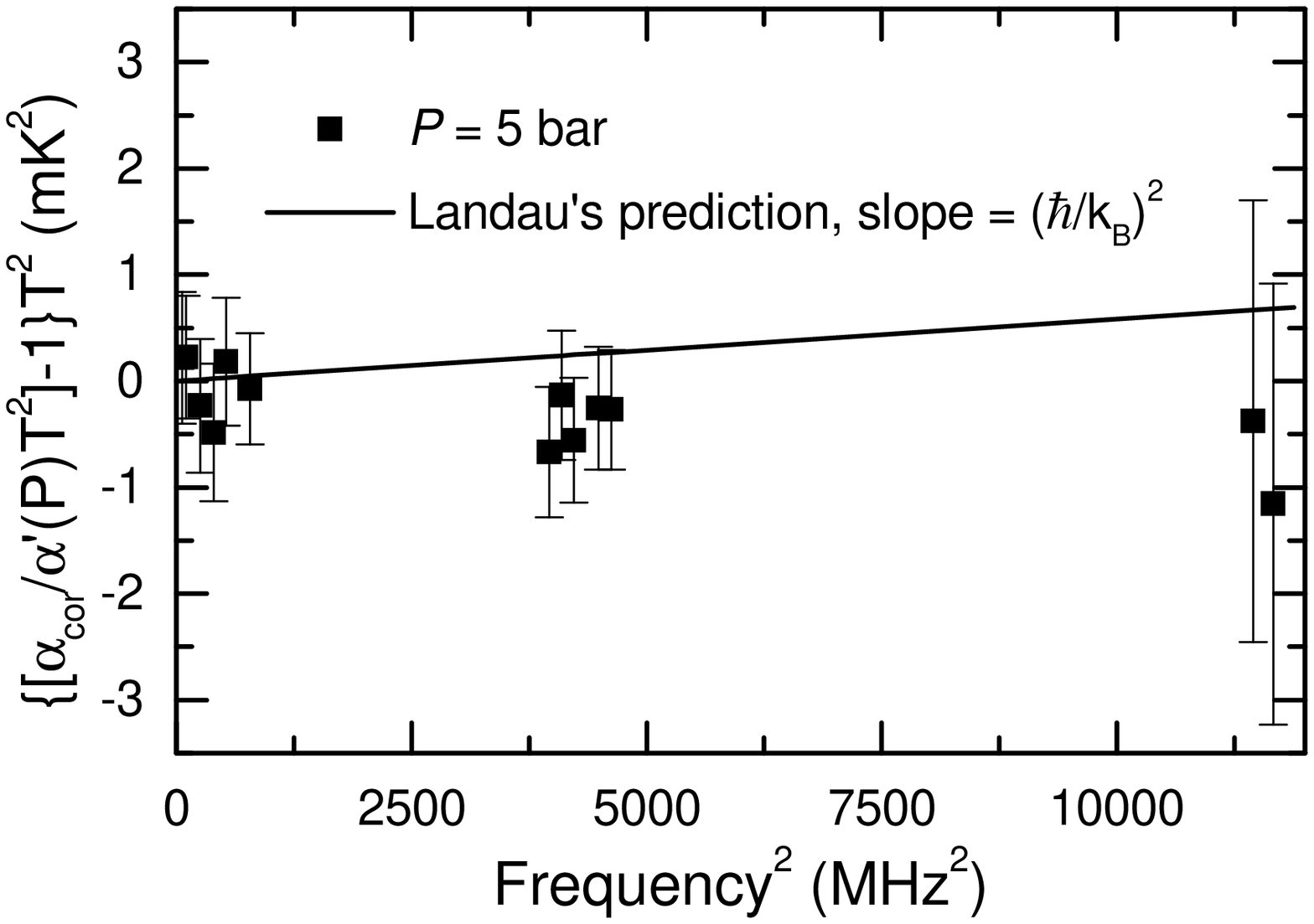}}
\caption{The corrected and normalized attenuation at 5 bar plotted 
as a function of the square of the frequency, Eq.~(1).  The data 
yield a slope of $(0.0 \pm 1.4)(\hbar/k_{B})^{2}$.}  
\label{fig:8}
\end{figure}

\begin{figure}
\centerline{\includegraphics[height=6.0in]{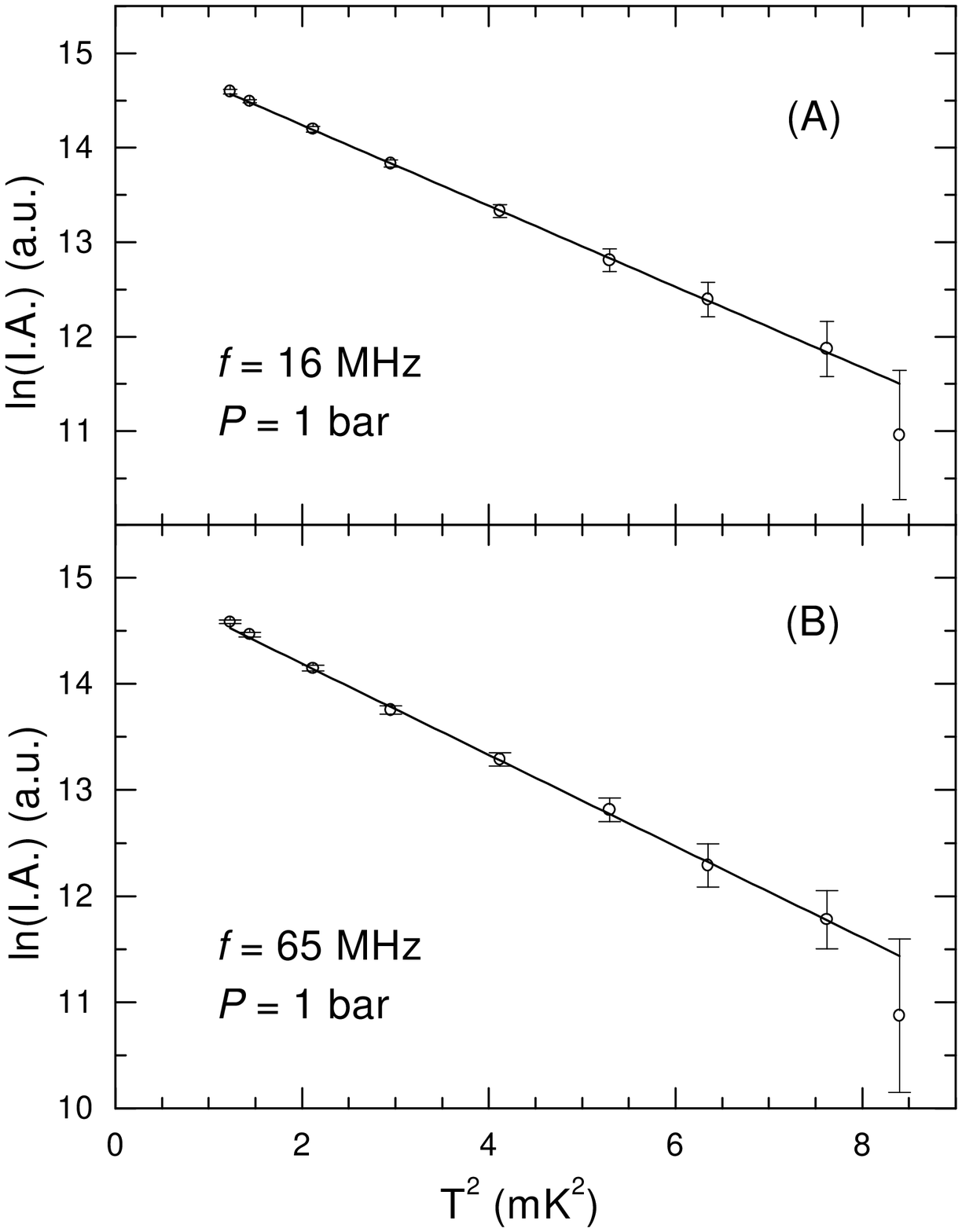}}
\caption{Typical data for the natural logarithm of the integrated amplitude (I.A.)  
in arbitrary units as a function of $T^2$ at 1 bar.}  
\label{fig:9}
\end{figure}

\begin{figure}
\centerline{\includegraphics[height=6.0in]{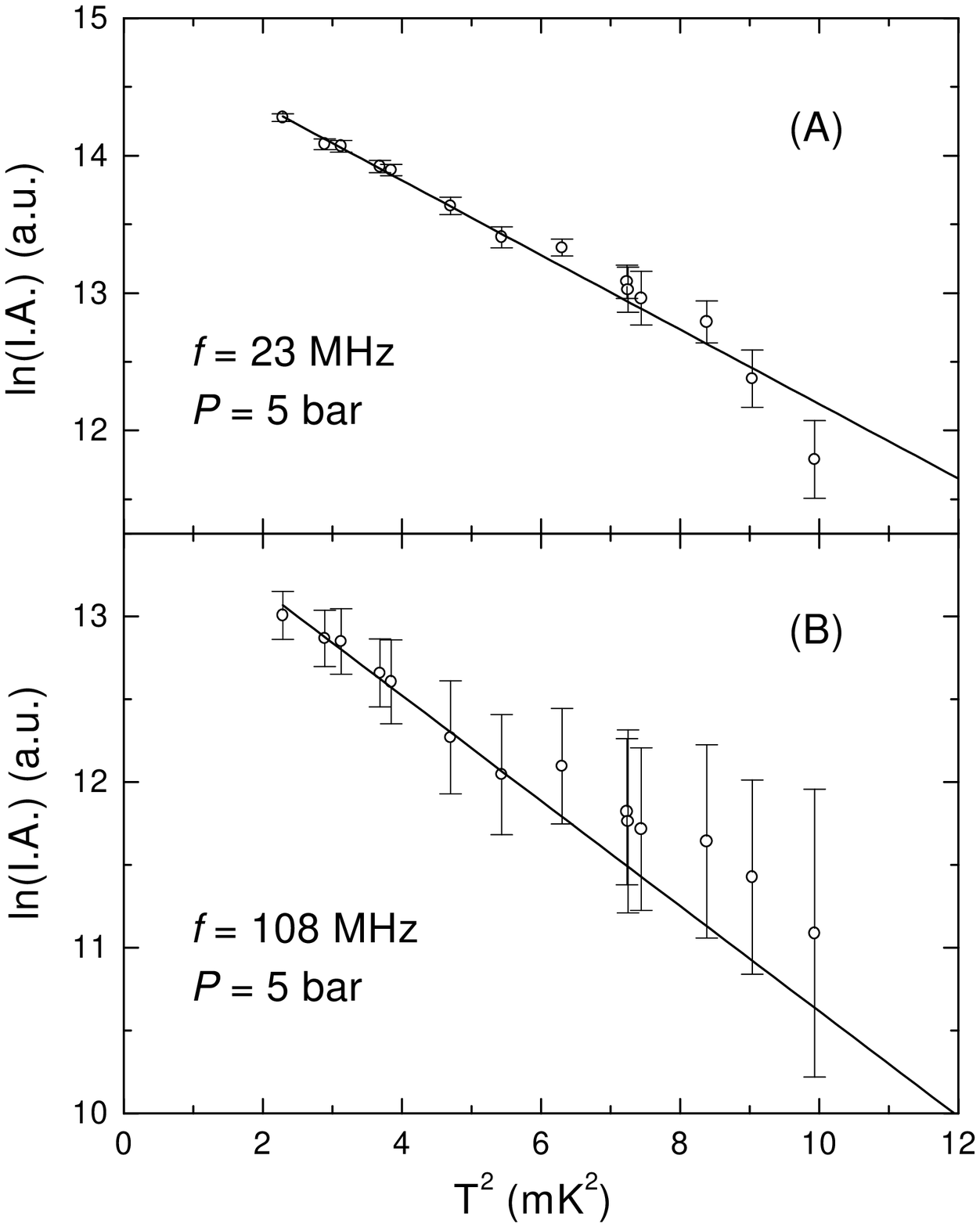}}
\caption{Typical data for the natural logarithm of the integrated amplitude (I.A.)  
in arbitrary units as a function of $T^2$ at 5 bar.}  
\label{fig:10}
\end{figure}

\begin{figure}
\centerline{\includegraphics[width=4.0in, height=3.0in]{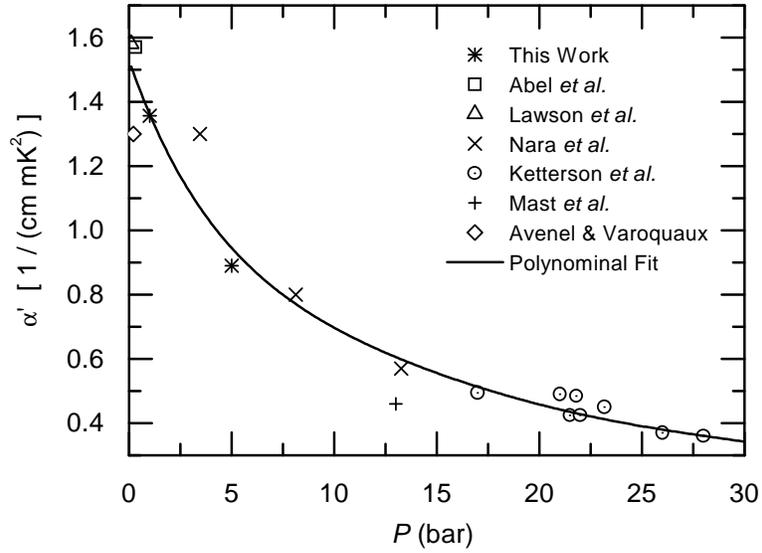}}
\caption{The values of $\alpha^{\prime}$, Eq.~(1), as measured by our work and 
other researchers.\cite{Abel2,Lawson,Ketterson,Mast,Nara,Avenel}  The solid line 
is from Eq.~(A3), see text.}  
\label{fig:11}
\end{figure}

\begin{figure}
\centerline{\includegraphics[width=4.0in,height=3.0in]{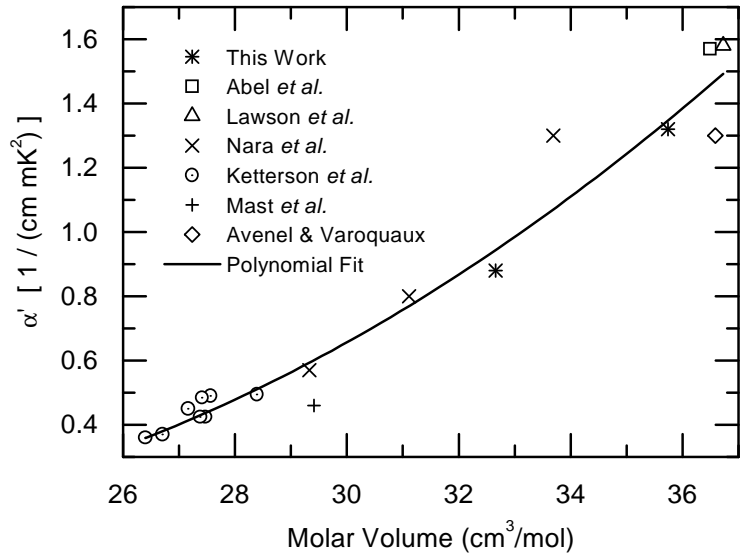}}
\caption{The values of $\alpha^{\prime}(P)$ shown in Fig.~11 have been 
converted to a function of molar volume using known values of 
$P(\nu)$.\cite{HV}  The solid line is from Eq.~(A2) which is a fit of the data.}  
\label{fig:12}
\end{figure}

\begin{figure}[p]
\centerline{\includegraphics[height=5.5in]{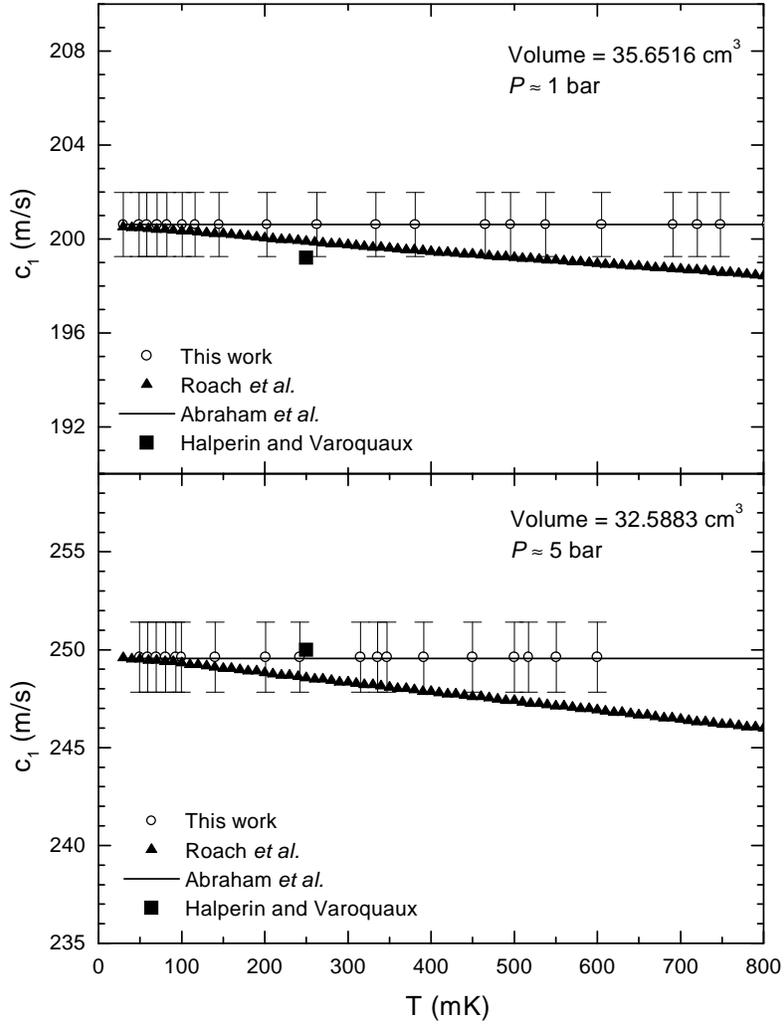}}
\caption{The velocity of first sound as a function of temperature at a pressure 
of 1 bar and 5 bar calculated using the path length 
and the measured signal delay for a 23 MHz pulse.  
The results of Roach \emph{et al.}\cite{Roach} and 
Abraham \emph{et al.}\cite{Abraham2} require input that was obtained 
from Kollar and Vollhardt,\cite{Kollar} and
also involve smaller shifts to normalize all the data to the values at 50 mK, 
see text.  The values as tabulated by Halperin and Varoquaux\cite{HV} do 
not involve any additional input parameters or shifts.}  
\label{fig:13}
\end{figure}  

\end{document}